# BRTSim, a general-purpose computational solver for hydrological, biogeochemical, and ecosystem dynamics

Federico Maggi

*Laboratory for Environmental Engineering, School of Civil Engineering, The University of Sydney, Bld. J05, 2006 Sydney, NSW, Australia.*

**Abstract**

This paper introduces the recent release v3.1a of BRTSim (BioReactive Transport Simulator), a general-purpose multiphase and multi-species liquid, gas and heat flow solver for reaction-advection-dispersion processes in porous and non-porous media with application in hydrology and biogeochemistry. Within the philosophy of the BRTSim platform, the user can define (1) arbitrary chemical and biological species; (2) arbitrary chemical and biological reactions; (3) arbitrary equilibrium reactions; and (4) combine solvers for phases and heat flows as well as for specialized biological processes such as bioclogging and chemotaxis. These capabilities complement a suite of processes and process-feedback solvers not currently available in other general-purpose codes. Along with the flexibility to design arbitrarily complex reaction networks, and setup and synchronize solvers through one input text file, BRTSim can communicate with third-party software with ease. Here, four cases study that combine experimental observations and modeling with BRTSim are reported: (i) water table dynamics in a heterogeneous aquifer for variable hydrometeorological conditions; (ii) soil biological clogging by cells and exopolymers; (iii) biotic degradation and isotopic fractionation of nitrate; and (iv) dispersion and biodegradation of atrazine herbicide in agricultural crops.

* Federico Maggi: federico.maggi@sydney.edu.au

# 1 Introduction

The increasing consensus that environmental research is fundamentally cross-disciplinary implies an increasing need to incorporate disciplines spanning across engineering, sciences, and computation. For example, while environmental engineering has largely focused on hydrological and geophysical water flows in aquifers and surface catchments, microbiology and chemical sciences have mainly focused on biochemical processes. Even if interactions across those disciplines have been limited, successful cases of cross-disciplinary approaches have eventually raised the attention of the scientific community to aspects such as global climate change, soil and water quality, and food security among the most important. The ground upon which synergistic research can converge and become effective is the coupling of process-based models that emphasise on hydrological and biogeochemical processes. Since the 80s, radical advances have been made in mechanistic accounting of physical processes and translation into computational frameworks. Within the hydrological sphere, physical processes of water flow through and at the soil surface are now recognized as important drivers for biological processes such as plant dynamics and microbial metabolism related to nutrients cycles and chemicals degradation, which eventually have important feedbacks on water flows. Examples of how these processes are now recognized to be highly coupled are available since a few decades, and a large number of research-oriented solvers has been produced. For example, there are likely hundreds of solvers for the water flow in soil that use the Darcy's law or integrate the Richards equation, and hundreds of codes for accounting ecohydrological fluxes linked to plant water and nutrient uptake. Manzoni and Porporato (2009) quote about 250 models for the coupled N and C cycles, and Heinen (2006) reports on 50 simplified models for nitrogen dynamics and losses. However, limited accessibility, obsolescence of coding environments, specialization of scripts, and incompatibility between platforms has hampered cross disciplinary computational capabilities, and discipline integration into shared computational environments is still limited even if potential applications are enormous in the modern era of information technology and attention to environmental quality.

Computational models for environmental applications would ideally encompass all suites of physical, chemical, and biological processes that govern mass and energy flows within and through the boundaries of an ecosystem in as mechanistic way as possible. This ideal scenario cannot be achieved as yet but the capabilities of mechanistic models to couple networks of processes using first principles have seen a steep development in recent years. As a consequence, gaps in process understanding and model capabilities or algorithmic robustness must be accompanied by uncertainty analysis at various levels including parameters and model structure. Accounting for a large number of coupled physical, chemical and biological processes involves the simultaneous solution of nonlinear partial differential equations used to describe individual processes, or the incorporation of scaling laws that approximate between-process dependencies when these are not mechanistically characterized. With a variable level of mechanistic and empirical coupling, and with obvious differences in computational structure and performance, successful solvers for hydrological applications include SHE/MIKESHE (Abbott et al., 1986), TOPMODEL (BevenFreer, 2001), VIC (Liang et al., 1994), and HYDRUS (Yu and Zheng, 2010) among others, while known models for coupled hydrological and inhert or reactive tracers for wider environmental applications include MODFLOW-RT3D (Johnson and Truex, 2006), TOUGHREACT (Xu et al., 2011), CENTURY (Parton et al., 1996) CASA (Potter et al., 1997), DNDC (Li et al., 1992), DAYCENT (Del Grosso et al., 2000), SWAT (Neitsch et al., 2000), RothC (Coleman and Jenkinson, 1996), MUSIC (eWater, 2018) and some of their derivatives. The list above is not exhaustive and a larger number of models for environmental applications can be found in various repositories including the CSDMS (Community Surface Dynamics Modeling System) and others. Besides differences in application domains and number of state variables solved for, the feature that mostly differentiates these solvers in two groups is their general-purpose capability to define a computational problem - not much in terms of space dimensionality but in terms of network of processes. For example, while the majority do not allow to define chemical species and design arbitrary reaction networks, or have only simplistic capabilities to do so, only TOUGHREACT has advanced capabilities to implement user-defined reaction networks. The limit that TOUGHREACT and other models have, however, is that the microbiological characterization is currently coarse and not yet developed to an advanced level. On the other hand, chemical kinetic solvers are available (e.g., ReKinSim Gharasoo et al. (2017), and KinTek Explorer Johnson et al. (2017)), but they do not include space dimensionality, and therefore cannot solve for space-time advection-dispersion-reaction environmental problems. In addition to those differences, only some models have interface capabilities that allow for interaction with data sets either for parameter estimation, or for model validation, and most of the times require recoding or coding of corollary scripts to conduct numerical analyses on uncertainty.

The aim of this work is to introduce the potentialities provided by the BRTSim (BioReactive Transport Simulator) computational framework as an analytical tool for hypothesis testing and parameter estimation, or as a predictive and scenario analysis tool in hydrology, biogeochemistry, and the wider environmental engineering and science. BRTSim is a general-purpose multi-phase and multi-species computational solver for biogeochemical reaction-advection-dispersion processes in non-isothermal porous and non-porous media. BRTSim can include user-defined ecohydrological boundary conditions and biogeochemical reaction networks, and finds suitable applications to describe water flow in soils and geophysical media, to track the transport and dispersion of aqueous and gaseous chemicals, as well as assess their chemical equilibrium and their decomposition rates in both chemical and biochemical reactions. More importantly, BRTSim allows for user-friendly management of model structure, and parameter estimation and validation thanks to built-in capabilities to run parallel simulation sessions. This work reports on existing and new capabilities implemented in the BRTSim v3.1a release, and shows application to four cases where modeling is compared to experiments (Case 1 and 3) or where BRTSim is used for analysis of numerical scenarios (Case 2 and 4, the latter including also experimental data). Although all cases are only exemplifications, they all are of interest to applications in hydrology and biogeochemistry and are intended to demonstrate greater potentialities such as those in Case 4 for the comprehensive description and analysis of contaminant dispersion an biodegradation in a cropland.

# 2 Material and methods

## 2.1 *Modeling founding concepts*

BRTSim solves for a user-defined multi-phase and multi-species system over an arbitrary physical domain made of grid elements that can have heterogeneous physical and hydraulic properties. The space dimensionality ranges from 0-D when one element is used, to 1-D when multiple elements are sequentially connected, or 1.5-D when multiple sequences of connected elements are used in a 2-D wise domain, with flows solved only through connected elements. BRTSim allows for the definition of any combination of single or multiple boundary flows in any element, be these constant or variable. BRTSim also allows defining a biogeochemical system with an arbitrary number of primary aqueous and biological species (PRI and BIO), and secondary aqueous (SEC), mineral (MIN) and gaseous (GAS) species. All secondary species are in chemical equilibrium with primary species. Species can be any real or virtual molecule, ion, substance or material; this capability allows the user to construct a system of tracers such as, for instance, to monitor the mass flow through a reaction and check mass balances with ease. The BIO species can be defined with any specific metabolic requirement and include properties of flow such as chemotaxis driven by chemical attractants and repellents in addition to diffusion and advection. Chemical and biochemical reactions can be defined for any kinetic order including multiple Michaelis-Menten-Monod terms, multiple competitive substrate consumption effects, and multiple inhibition terms. Finally, BRTSim v3.1a accounts also for heat flow and temperature effects on physical properties of phases, and on activity of BIO species according to user-defined temperature sensitivity parameters.

## 2.2 *Computational principles*

BRTSim solves for the mass continuity and conservation laws of the phases (liquid L, gaseous G, biological B, and mineral M), and chemical and biological species within those phases. The mass $M_\beta$ of phase $\beta$ in an element of volume $V$ of the solving domain is $M_\beta = \phi S_\beta \rho_{\beta e} V$ for $\beta =$ L, G and B, and $M_\beta = (1-\phi)\rho_\beta V$ for $\beta =$ M. Neglecting dynamic changes in phase M, the mass conservation law for the other phases can be written in a general form as

$$\frac{\partial M_\beta}{\partial t} = \int_\Gamma \rho_{\beta e} v_\beta \mathrm{d}\Gamma + \int_V \rho_{\beta e} u_\beta \mathrm{d}V \pm \int_V f_\beta r_B \rho_{\beta e} \mathrm{d}V, \qquad \text{for} \quad \beta = \text{L,G,B}. \tag{1}$$

The first term on the RHS of Eq. (1) represents the internal fluxes where $v_\beta$ corresponds to the velocity of phase $\beta$ through the surface $\Gamma$ of two adjacent volumes, be this expressed by the Darcy's equation for $\beta =$ L,G or transport and chemotaxis for $\beta =$ B. The second term includes the rate $u_\beta$ of production or destruction of phase $\beta$ in $V$ and practically represents the source or sink of mass through the boundary of each element from outside the solving domain in contrast to the mass fluxes across other volumes of the solving domain. The third term in Eq. (1) is an internal mass exchange with rate $r_\beta$ between phases and currently exists only for the L and B phases, with $f_\beta$ the coefficient of conversion. In specific, BRTSim accounts for immobilization and remobilization of the L phase into the B phase as a result of the dynamics of all declared BIO species after kinetic growth or mortality. Hence, $f_\beta = 0$ for the G phase, or when only the L phase exists, or $f_\beta \neq 0$ for the L and B phases if BIO species exist and undergo kinetic processes.

The velocity $v_\beta$ in Eq. (1) is written as

$$v_\beta = -k \frac{k_{r\beta}(S_\beta)}{\mu_\beta} \nabla [P_\beta(S_\beta) + \rho_{\beta e} g z] \qquad \text{for} \quad \beta = \text{L,G}, \qquad (2a)$$

$$v_\beta = \epsilon_\beta v_L X_\beta - D_\beta \nabla X_\beta + D_{\beta,c} X_\beta \nabla X_{L,c} - D_{\beta,r} X_\beta \nabla X_{L,r} \quad \text{for} \quad \beta = \text{B}. \qquad (2b)$$

Eq. (2)a is essentially the Darcy's velocity and includes the absolute and relative permeability $k$ and $k_{r\beta}$, respectively, the gravitational acceleration $g$, the total pressure $P_\beta$ of phase $\beta$ (including the osmotic pressure caused by dissolved species in the L phase), and the vertical position $z$ relative to the system of reference. The relative permeability $k_{r\beta}(S_\beta)$ and pressure $P(S_\beta)$ are calculated as a function of the phase saturation $S_\beta$ either with the Brooks and Corey (1962) or the van Genuchten (1990) model.

Eq. (2)b only applies to phase $\beta = \text{B}$, where $X_\beta$ is the mass fraction of BIO species $X$; the first term on the RHS expresses the transport of $X_\beta$ proportionally to an efficiency $\epsilon_\beta$ describing the detachment fraction of BIO species and the Darcy's velocity in the L phase. The second term expresses the diffusion of BIO species proportionally to the diffusion coefficient $D_\beta$. The third term expresses chemotaxis of BIO species with a modified Keller and Segel (1971) model against the concentration gradient of chemical attractant $X_{L,c}$ in phase L with coefficient $D_{\beta,c}$, while the fourth term expresses chemotaxis away from the concentration gradient of chemical repellents $X_{L,r}$ in phase L with coefficient $D_{\beta,r}$.

The total mass $M_\beta^k$ of species $k$ in phases $\beta = \text{L,G}$ in a volume $V$ of the solving domain is $M_\beta^k = \phi S_\beta \rho_{\beta e} V X_\beta^k$, where $X_\beta^k$ is the mass fraction of species $k$ in $\beta$. The mass conservation law for $M_\beta^k$ can be written similarly to Eq. (1) as

$$\frac{\partial M_\beta^k}{\partial t} = \int_\Gamma \rho_{\beta e} (v_\beta X_\beta^k - D_\beta^k \nabla_z X_\beta^k) d\Gamma + \int_V \rho_{\beta e} u_\beta X_\beta^k dV \pm \int_V r_\beta^k \rho_{\beta e} dV. \qquad (3)$$

The first term on the RHS is the internal mass flux of species $k$ in phase $\beta$ through the surface $\Gamma$ of a volume in the solving domain by Darcy's advection as in Eq. (2) and molecular diffusion, with $D_\beta^k$ the diffusivity of species $k$ in phase $\beta$. The second term is the integral of all mass sources or sinks of species $k$ within phase $\beta$. The third term is the integral of all mass exchange rate of species $k$ in the two phases $\beta =$ L,G, with $r_\beta^k$ the rate of production or destruction of species $k$ resulting from both kinetic and equilibrium reactions.

Heat flow is explicitly described by means of conduction through the phase $\beta$ across elements of the solving domain and across phases within the same element, and by advection of phases L and G. Details are not reported here as this capability is not directly used in the examples presented in this work (see formulations in the Technical Guide and User Manual of BRTSim available at the link in Appendix).

Kinetic reactions are assumed to occur only in the L phase and only apply to primary PRI and BIO species. The rate of change $r_\beta^k$ of species $k$ in phase $\beta=$ L introduced in Eq. (3) can be expressed as

$$r^k = \frac{\mathrm{d}X^k}{\mathrm{d}t} = x_k R, \tag{4}$$

where $x_k$ is the stoichiometric number for component $k$ and $R$ is the reaction velocity. Kinetic reactions can be defined for any kinetic order and rate constant, and with an arbitrary number of n-th order kinetic products, Michaelis-Menten-Monod terms (Michaelis and Menten, 1913; Monod, 1949), competition terms, and inhibition terms. For a generic reaction with $n_O$ n-order kinetic products (either PRI or BIO species, or both), $n_{MM}$ Michaelis-Menten-Monod terms, $n_{COM}$ competitive reactants, and $n_{INB}$ inhibition terms, the reaction velocity $R$ is written as

$$R = r f_B \prod_{n_O} X_{n_O}^{x_{n_O}} \cdot \prod_{n_{MM}} \frac{X_{n_{MM}}}{X_{n_{MM}} + K_{n_{MM}} \left(1 + \sum_{n_{COM}} \frac{X_{n_{COM}}}{K_{n_{COM}}}\right)} \prod_{n_{INB}} \frac{X_{n_{INB}}}{X_{n_{INB}} + K_{n_{INB}}}, \tag{5}$$

where $r$ is the reaction rate constant, and $f_B = 1$ if no BIO species are defined in that reaction. If BIO species are used, then $f_B \leq 1$ is defined as

$$f_B = \min\{f(S_B), f(T), f(S_L)/\max\{f(S_L)\}\}, \tag{6}$$

with $f(S_B)$, $f(T)$ and $f(S_L)$ the specific microbial response functions defined as

$$f(S_B) = \min\left\{1 - \frac{S_B - S_{Lr}}{1 - S_{Lr} - S_{Gr}}, 1 - \frac{f_L S_B}{S_L}, 1 - \frac{(1 - f_L)S_B}{S_G}\right\}, \tag{7a}$$

$$f(T) = \frac{e^T}{e^{T_{LB}} + e^T} \cdot \frac{e^{T_{UB}}}{e^{T_{UB}} + e^T}, \tag{7b}$$

$$f(S_L) = \frac{S_L}{S_{L,LB} + S_L} \cdot \frac{S_{L,UB}}{S_{L,UB} + S_L}, \tag{7c}$$

where $T_{LB}$ and $T_{UB}$, and $S_{L,LB}$ and $S_{L,UB}$ are the lower and upper temperature (in Kelvin) and liquid saturation response parameters that can be benchmarked against data in Yan et al. (2018); Feller (2010); Skopp et al. (1990). A different approach to account for temperature effects on biological reaction will be used in future releases based on the thermodynamic approach recently developed in Maggi *et al.* (2018), and is not explored in further detail in this work. The function $f(S_B)$ implies that BIO species can grow as long as there is enough free L phase to immobilize, or G space available for the solid fraction of BIO species to occupy, or there is enough pore volume to host the total BIO volume. Function $f(T)$ limits $R$ when $T$ is below $T_{LB}$ and above $T_{UB}$, while function $f(S_L)$ limits $R$ when $S_L$ is below $S_{L,LB}$ or above $S_{L,UB}$.

Finally, equilibrium reactions are solved for the mass fraction or partial pressure of all SEC species as a function of PRI species. Equilibrium reactions are defined by an equilibrium constant $K_{eq}$ at a given temperature $T$, and are defined for species in the L, M and G phases according to the mass-action law

$$K_{eq}(T) = X_\beta^{x_k} \prod_j X_j^{x_j}, \qquad (8)$$

where $X_\beta$ is the a secondary species in phase $\beta$ that is to be solved for with $x_k$ its stoichiometric number, while $X_j$ and $x_j$ are the list of PRI species and their stoichiometric numbers in that equilibrium reaction, respectively.

Discretization of Eq. (1), Eq. (3), Eq. (4) and Eq. (8) is not described here, but full detail on the finite volume explicit and implicit methods can be accessed in the Technical Guide and User Manual of BRTSim available at the link in Appendix.

*2.3 File management and solvers synchronization*

BRTSim requires the `Param.inp` file, which contains all TXT information needed to solve a computational problem, and produces the `Grid.out` file listing all material properties of the computational problem, the `Time.out` file containing the state variables calculated in each node and at each or selected time steps, and the `Flux.out` file with all cumulative fluxes of phases, species and heat through the boundaries at each or selected time steps. The optional `TimePST.out` file may be generated for selected compounds, grid nodes, and times (Figure 1a).

BRTSim includes solvers for (1) the L flow, (2) the G flow, (3) heat flow, (4) transport of species (TR), (5) chemotaxis (CTX), (6) kinetic reactions (KIN); and (7) equilibrium reactions (EQ) (Figure 1b). These solvers can be switched on/off depending on the problem to be solved, but the L phase solver is by default on. Solvers are synchronized to run sequentially but convergence is controlled on each individual solver through the `Param.inp` file with criteria that typically are process dependent. If convergence is not reached in one of the TR, CTX, KIN and EQ solvers, these are re-iterated until when the solved $\Delta t$ equals $\Delta t$ of the L phase flow solver (Figure 1b). This assures that each process is homogeneously solved in time. Mass balance is checked in each iteration and the user can verify how solution meets user-defined convergence criteria.

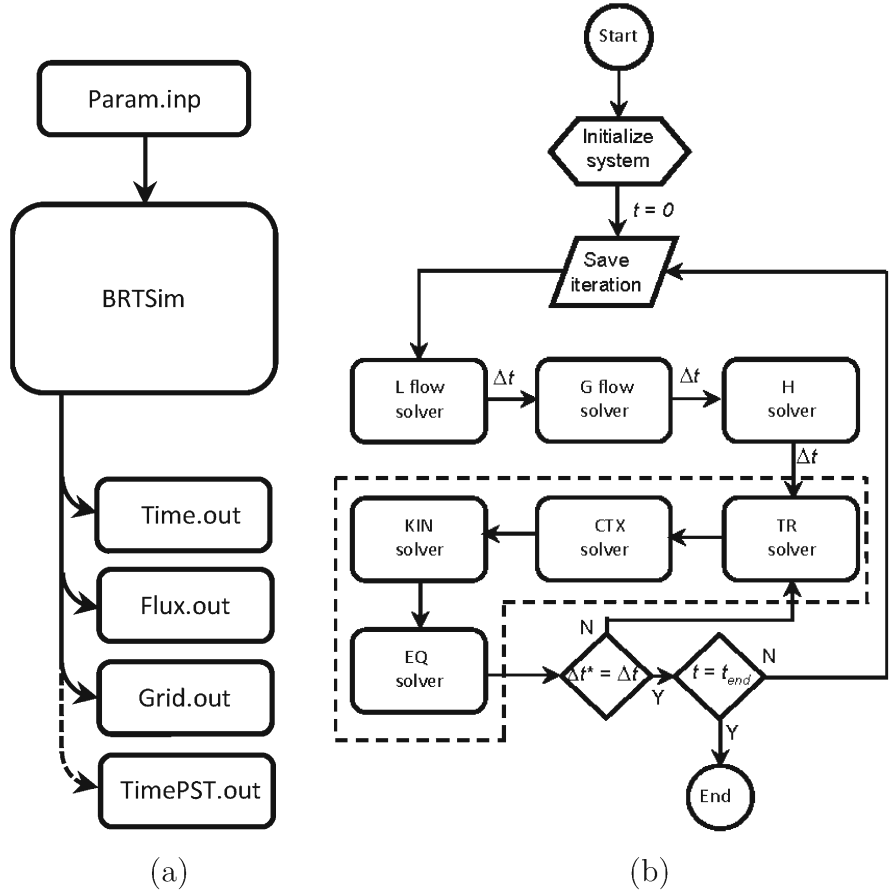

(a)   (b)

Fig. 1. (a) File organization and (b) sequence of solvers calls and synchronization in BRTSim

*2.4 Examples of generalized use of BRTSim*

To demonstrate application of BRTSim, four cases are briefly presented, each emphasising a specific hydrological or biogeochemical process, or a combination of the two.

The first example (Case 1) is the modeling and comparison with observations of the water table dynamics in an heterogeneous aquifer subject to variable hydrometeorological boundary flows, irrigation, and crop evapotranspiration. In this example, experimental hydraulic parameters of the soil layers that characterize the aquifer were used with deterministic simulation to predict the water table dynamics, while a stochastic analysis was used to highlight how variability in permeability, air-entry suction and pore volume distribution index affected the water table depth over time and how this compares with observations.

The second example (Case 2) shows the effect of biological clogging of a soil column caused by production of exopolymeric substances ($EPS$) by microorganisms that immobilize water and reduce the hydraulic conductivity. This example shows how modeling can be used for *a priori* analyses of a coupled hydraulic and biogeochemical problem that can find application in contaminant hydrology and bioremediation engineering.

The third example (Case 3) refers to nitrate reduction and its isotopic fractionation by a soil microcosm in laboratory controlled conditions. A combinatorial isotopic assemblage is used to demonstrate application of recent advances in isotopic tracking and modeling, and is accompanied by a sensitivity analysis to microbial response to temperature.

The last example (Case 4) is a brief summary of the modeling of dispersion and biodegradation of atrazine, and includes full degradation reaction networks and a comprehensive suite of ecohydrological processes and boundary conditions applied to an agricultural soil profile. This example is not presented in full, but only key finding are introduced to stimulate reasoning about potential large scale applications of this or similar approaches after earlier analysis of parametric and model uncertainty described later.

## 3 Results

*3.1 Case 1 - Water table dynamics in an heterogeneous aquifer*

This first example proposes the modeling of the water table dynamics and comparison with observations from a monitoring well located near Reggio Emilia, Italy (40°45′22″N; 10°41′5″E). The aquifer is characterized by a shallow water table (less then 5 m) and heterogeneous textural composition to a depth of 5 m resulting from alluvial sedimentary deposits from the nearby Po river. The textural properties were retrieved from the (SGSS, 2016) regional geological repository, and were used to estimate the mineral density $\rho_m$, porosity $\phi$, air-entry suction $\psi_s$, pore size distribution index $b$, and absolute permeability $k$ using Cosby et al. (1984) (Table 1). Soil is cultivated with vines for wine production. With an assumed 20% canopy interception, the vineyard is exposed to an average effective precipitation $P_I = 540$ mm/year (Arpae-Simc), and potential evapotranspiration $ET_0 = 1133$ mm/year. The crop evapotranspiration $ET_c = K_c \times ET_0$ was calculated from Allen et al. (1998) with the crop coefficient $K_c$ varying during the growing season. Irrigation timing and rates were estimated from Chiari et al. (2016). Observations of water table depth extended from 2006 to 2016 with an average frequency of about two weeks.

BRTSim was used to replicate the water table dynamics with the accounting of ecohydrological boundary flows and soil hydraulic characteristics. The environmental setting was suitable to use BRTSim with a full 1-D domain extending from the soil surface to 5 m depth; soil included 4 layers with various hydraulic properties, which were described with 52 grid elements with the upper being the atmosphere; daily precipitations and irrigations were applied to the second node from the top, which represented the soil surface. In contrast, crop evapotranspiration was accounted for by water uptake down to 1.4 m with a distribution that replicated the typical vine root system.

| [a]Depth (m) | [b]Soil texture | [a]Sand-Silt-Clay (%-%-%) | [c]$\rho_m$ (kg m$^{-3}$) | [c]$\phi$ (-) | [c]$b$ (-) | [c]$\psi_s$ (m) | [c]$k$ (m$^2$) × 10$^{-13}$ |
|---|---|---|---|---|---|---|---|
| 0-0.5 | Silt loam | 16-60-24 | 2848 | 0.469 | 6.73 | -0.47 | 1.66 |
| 0.5-0.8 | Silt loam | 9-66-25 | 2855 | 0.478 | 6.88 | -0.58 | 1.29 |
| 0.8-1.2 | Silt loam | 12-73-15 | 2852 | 0.474 | 5.29 | -0.53 | 1.44 |
| 1.2-5.0 | Silt loam | 12-68-20 | 2852 | 0.474 | 6.09 | -0.53 | 1.44 |

Table 1
Summary of soil parameters. [a] From SGSS (2016). [b] USDA soil texture classification system. [c] Estimated using Cosby et al. (1984).

Results show that observed and estimated boundary water fluxes caused the water table to fluctuate mainly through the top 3 m of soil, and was replicated relatively well over the period ranging from 2006 to 2016 (Figure 2), with a correlation R = 0.82 and normalized root mean square error NRMSE = 18.6%. At times, the predicted water table depth departed from observations likely because of the use of a one dimensional mesh (i.e., lateral flows were not accounted for), or because of an heterogeneous the vine root systems or soil structure. The latter instance is a typical case for application of stochastic sensitivity analysis; specifically, the hydraulic parameters of interest (i.e., $k$, $\psi_s$ and $b$) were cloned in 50 replicas with values randomly extracted from a Guassian distribution function with average equal to experimental values reported in Table 1 and standard deviation equal to 50% of the experimental value. The 50 values of each parameters were independently used in an equal number of model simulations, which were compared with each other to highlight which parameter had the most important effect against observations. Results in Figure 3 show that the permeability $k$ only affected slightly the water table fluctuations, while a larger effects were found relative to the air-entry suction $\psi_s$ and pore volume distribution index $b$. Variability in model output better captured experimental observations, and it is therefore suggested that assessment of uncertainty can become useful to identify model sensitivity and parameters that require better experimental characterization for modeling purposes.

*3.2 Case 2 - Soil biological clogging*

Earlier and recent works (e.g., Maggi and Porporato, 2007; Brangari et al., 2018) have shown that microbial film formation from exopolymeric substances (EPS) and cell accumulation can lead to local biological clogging of pore networks in soil and affect the soil hydraulic characteristics and drainage. In this second example, biological clogging and its effect on water infiltration in a vertical soil column is illustrated for applications in bioremediation strategies such as biobarriers to contaminants (e.g., Ross et al., 1997) or biological systems that integrate barriers with biodegradation (e.g., Cunningham et al., 2003). Specifically to this exercise, modeling was used as a tool to test hypotheses of microbial EPS production rate and biological clogging effectiveness.

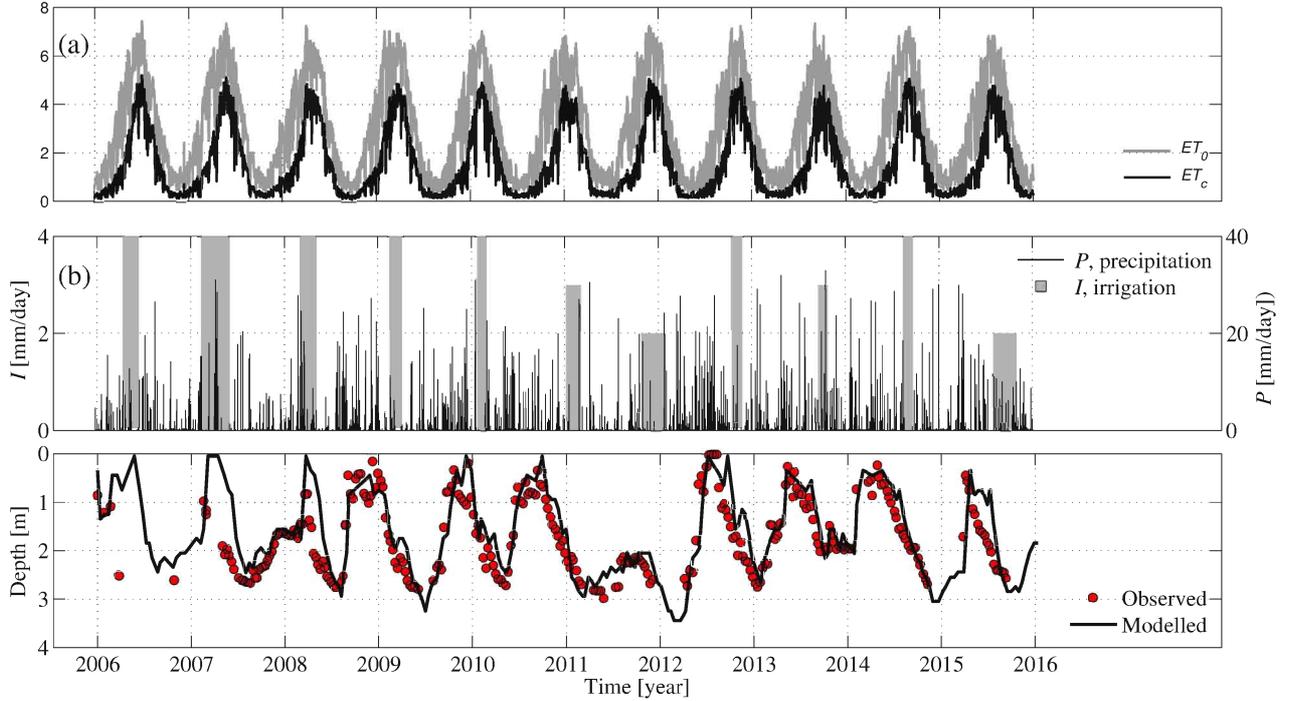

Fig. 2. (a) daily estimated potential and crop evapotranspiration; (b) daily observed precipitation and applied irrigation; and (c) observed and modelled water table depth in a vineyard over 12 years near Reggio Emilia, Italy.

A 1.6 m soil column of surface area $A = 1$ m$^2$ was subject to a water flow rate of $5 \times 10^{-7}$ m$^3$/s from time $t = 0$ for 2 days applied at the soil top, and free drainage at the bottom. Soil textural fractions were 90% sand, 5% silt and 5% clay with mineral bulk density of 2848 kg/m$^3$ and porosity $\phi = 0.46$. The absolute permeability $k = 2.24 \times 10^{-12}$ m$^2$, air-entry suction $\psi_s = -5.02 \times 10^{-2}$ m, and pore volume distribution index $b = 3.705$ of the Brooks and Corey (1962) water saturation-tension relationship were calculated with the Cosby et al. (1984) empirical laws. The soil column was initialized with a uniform water saturation $S_L = 0.55$, hence, the inflow produced a water front propagating downward over time. Microorganisms ($B_{EPS}$) located between 1 and 1.2 m depth at a concentration of $10^2$ mg/L (about 17 mg/kg$_{\text{dry soil}}$ assumed constant over time) produced exopolymeric substances ($EPS$) with a first-order kinetic rate $r$ of $B_{EPS}$ that caused clogging. The $EPS$ substance is defined in this example by mass density $\rho = 1025$ kg/m$^3$ and a water volume fraction 0.8 (up to 0.98 was observed in Yang and Li, 2009), and it is assumed not to undergo diffusion and advection. For simplicity in this illustrative example, $EPS$ production by $B_{EPS}$ was assumed not to be limited by nutrient availability - detailed microbiological nutrient metabolic requirement is illustrated in Case 4. $EPS$ underwent degradation over time with a first-order reaction rate $\delta$ of $EPS$ concentration, the process that remobilized water after $EPS$ lysis (Maggi and Porporato, 2007). The dynamics that govern $EPS$ concentration can therefore be described as

$$\frac{\mathrm{d}EPS}{\mathrm{d}t} = f_B r B_{EPS} - \delta EPS, \tag{9}$$

with the equilibrium $EPS$ concentration calculated analytically as $EPS_{eq} = f_B r B_{EPS}/\delta$. For this example, the response of $B_{EPS}$ to $S_L$ was defined in $f_B$ of Eq. (7) by $S_{L,LB} = 0.3$ and $S_{L,LB} = 0.8$, and the effect of porosity was explicitly accounted for as in Eq. (7)a, while the effect of temperature was neglected. Dynamic biological clogging was implemented in BRTSim by means of standard writing of 2 kinetic reactions for $EPS$ production and lysis in the `Param.inp` file.

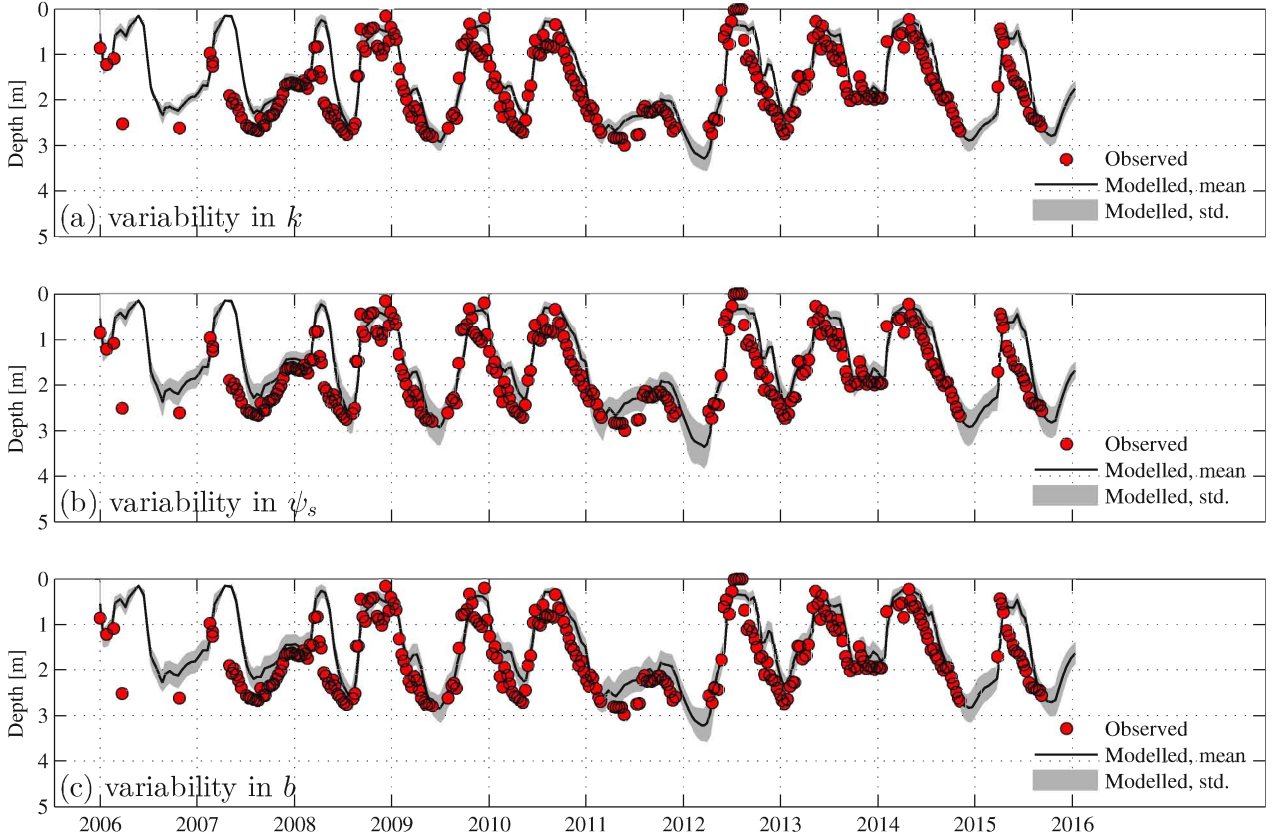

Fig. 3. Stochastic sensitivity analysis of the waer table fluctuations relative to: (a) soil permeability; (b) air-entry suction; and (c) pore volume distribution index. 50 replicas per each parameters were generated; black line represents the average while gray area represents the standard deviation of the set of simulations.

BRTSim was run with $r = 7.5 \times 10^{-2}$ 1/s and $\delta = 10^{-6}$ 1/s. The profiles in Figure 4 show the mobile water front propagation (panel $a$), the total water saturation (mobile and immobilized in *EPS*, panel *b*), and the concentrations of $B_{EPS}$ and *EPS* (panels *c* and *d*). The initial increase in *EPS* concentration led to water immobilization when the water front passed by. At this time, clogging substantially decreased the capability for the soil to conduct water beneath, and a water table formed above the clogged section. Changes in microbial activity due to available water in the clogged section produced slower *EPS* production at the centre of the clogged section, the mechanism that introduce nonlinearities in *EPS* content. Next, the front reached the bottom of the soil column draining at reduced rate as compare to above the clogged soil section.

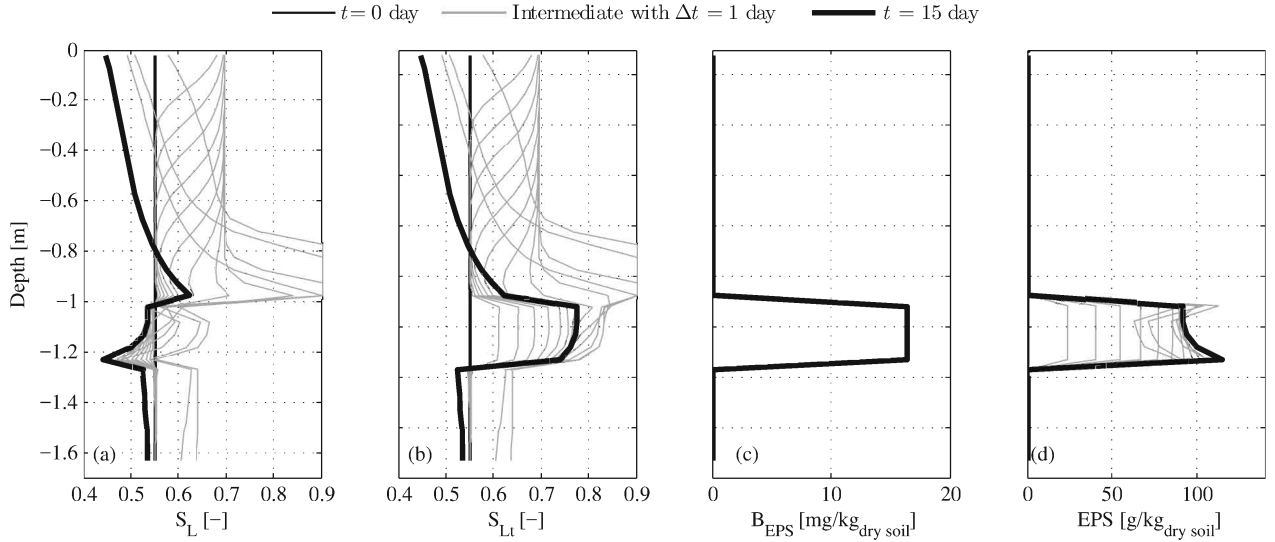

Fig. 4. Vertical profile of (a) mobile water saturation $S_L$; (b) total water saturation (mobile and immobile) $S_{Lt}$; (c) microbial biomass fraction $B_{EPS}$; and (d) *EPS* mass fraction. Profile are represented at time $t = 0$ day and at intervals $\Delta t = 1$ day up to final time $t = 15$ day.

*EPS* production rates $r$ ranging between 0 and 1 1/s were used to highlight differences in water flow against various clogging rates. Over time, maximum clogging did not completely stop water to drain through the clogged section in this case, but largely decreased the flow rate as compared to no clogging (Figure 5a). Intermediate clogging rates highlight two aspects: (i) the maximum flow rate at the bottom of the column occurred at a later time as compared to no clogging; and (ii) an initial reduction in flow rate from day 3 to day 5 was observed during water immobilization in *EPS* where clogging occurred (Figure 5a); this latter effect was caused by reduced permeability, accounted for in BRTSim using the scaling proposed in Brutsaert (2000) and Maggi and Porporato (2007) as $k' = k(1 - S_B)^2$, and by an increased suction at the clogged section, where water was immobilized into *EPS*. Intermediate bioclogging rates produced intermediate solutions (Figure 5a). The most important effect was noted in the highly nonlinear total *EPS* content (Figure 5b), which showed an oscillatory dynamics in the first 6 to 12 days caused by the water immobilized at and accumulating above the clogged section, and the following slow release. After the water front passed by, *EPS* production continued at a decreasing rate due to a lower water saturation. This dynamic feedback has not been observed yet in experimental studies and these result may be used as an indication for further research.

Different scenarios can be obtained when the soil wetness is subject to dynamic changes such as in field conditions.

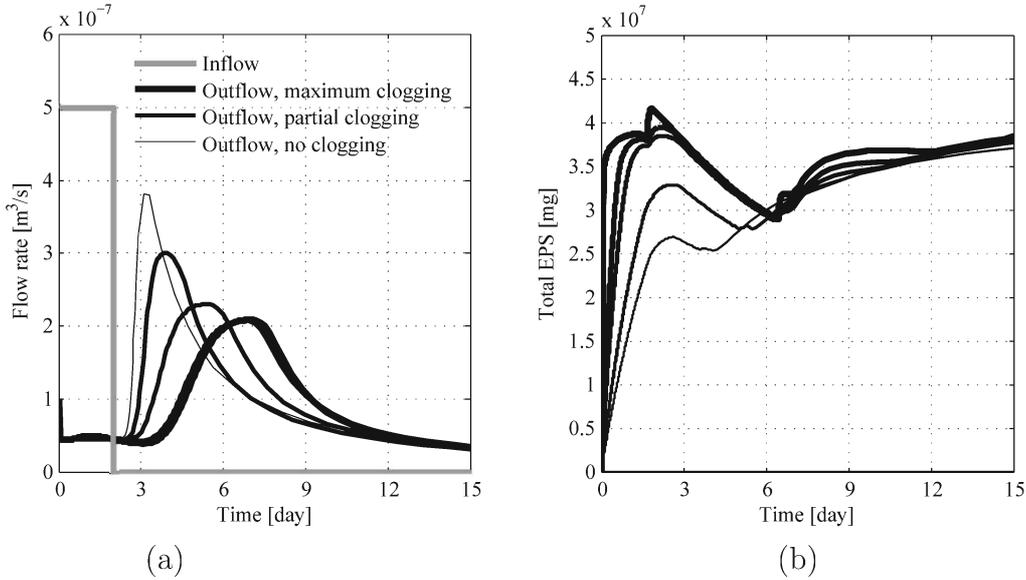

Fig. 5. (a) Inflow rate at the top soil and outflow rates at the bottom draining section as a function of time for various bioclogging regimes determined by $EPS$ production rates $r$; and (b) total $EPS$ mass fraction as a function of time for $EPS$ production rates $r$ ranging from $r = 0$ 1/s to 1 1/s.

### 3.3 Case 3 - Isotope fractionation of $NO_3^-$ denitrification

Stable isotopes are normally used in hydrological and biogeochemical applications to track sources and pathways of water and nutrients, but models that comprehensively account for isotopic speciation and fractionation by biological and physical processes are rare (e.g., Thullner etal., 2008), or only favour simplistic approaches that use the Rayleigh equation. In this third example, BRTSim was used to describe $NO_3^-$ denitrification and isotope fractionation in a 0-D description of a soil microcosm based on laboratory experiments conducted at 20 $^oC$ in Mariotti *et al.* (1981). The Generalized Equations for Biochemical Isotope Kinetics (GEBIK, Maggi and Riley, 2009, 2010) describe the combinatorial competitive denitrification pathways reducing $^{14}NO_3^-$ and $^{15}NO_3^-$ to $N_2O$ as

$$2\,^{14}NO_3^- \rightarrow\,^{14}N_2O, \tag{10a}$$
$$^{14}NO_3^- +\,^{15}NO_3^- \rightarrow\,^{14}N^{15}NO, \tag{10b}$$
$$^{14}NO_3^- +\,^{15}NO_3^- \rightarrow\,^{15}N^{14}NO, \tag{10c}$$
$$2\,^{15}NO_3^- \rightarrow\,^{15}N_2O. \tag{10d}$$

Excluding Eq. (10)d and combining Eq. (10)b and c, the reactions were rewritten as

$$2\,^{14}\text{NO}_3^- \rightarrow {}^{14}\text{N}_2\text{O}, \tag{11a}$$

$$^{14}\text{NO}_3^- + {}^{15}\text{NO}_3^- \rightarrow u\,^{14}\text{N}^{15}\text{NO} + (1-u)\,^{15}\text{N}^{14}\text{NO}, \tag{11b}$$

where Eq. (11)b describes both isotopomer N$_2$O product expressions with $u = 50.225$ % the partition coefficient for $^{14}$N$^{15}$NO and $(1-u) = 49.775\%$ for $^{15}$N$^{14}$NO (Well *et al.*, 2006). The GEBIK equations were used to describe the rates of change of $^{14}$NO$_3^-$, $^{15}$NO$_3^-$, and biomass dynamics under the quasi-steady-state assumption. With the hypothesis that the enzyme $E$ is produced and degrades at the same rate as cells grow and age (Maggi *et al.*, 2018), the GEBIK equations become

$$\frac{\text{d}[^{14}\text{NO}_3^-]}{\text{d}t} = -2\frac{zk_1[B][^{14}\text{NO}_3^-]}{[^{14}\text{NO}_3^-] + K_1\left(1 + \frac{[^{15}\text{NO}_3^-]}{K_2}\right)} - \frac{zk_2[B][^{15}\text{NO}_3^-]}{[^{15}\text{NO}_3^-] + K_2\left(1 + \frac{[^{14}\text{NO}_3^-]}{K_1}\right)}, \tag{12a}$$

$$\frac{\text{d}[^{15}\text{NO}_3^-]}{\text{d}t} = -\frac{zk_2[B][^{15}\text{NO}_3^-]}{[^{15}\text{NO}_3^-] + K_2\left(1 + \frac{[^{14}\text{NO}_3^-]}{K_1}\right)}, \tag{12b}$$

$$\frac{\text{d}[B]}{\text{d}t} = Y\left(-\frac{\text{d}[^{14}\text{NO}_3^-]}{\text{d}t} - \frac{\text{d}[^{15}\text{NO}_3^-]}{\text{d}t}\right) - \delta[B], \tag{12c}$$

where $z = 10^{-10}$ mol/mg is the enzyme yield coefficient after Maggi and La Cecilia (2016), $k_1$ and $k_2$ are the reaction rate constants, $K_1$ and $K_2$ are the Michaelis-Menten constants (Michaelis and Menten, 1913), $Y$ is the biomass yield coefficient, and $\delta = 10^{-6}$ 1/s is the biomass mortality rate. Next, the Generalized Equations for Biochemical Isotope Fractionation (GEBIF) were used to describe the time-dependent isotopic ratio for the substrate $R_S$ and the standard isotopic ratio in $\delta$‰ as (Mariotti *et al.*, 1981; Maggi and Riley, 2010)

$$R_S(t) = \frac{15[^{15}\text{NO}_3^-(t)]}{14[^{14}\text{NO}_3^-(t)]}, \tag{13a}$$

$$\delta^{15}\text{N-NO}_3^-(t) = \left(\frac{R_S(t)}{R_{std}} - 1\right) \times 1000, \tag{13b}$$

with $R_{std} = 0.0229$ the standard natural $^{15}$N abundance, and NO$_3^-(t)$ concentrations expressed in mol/L.

In this example, implementation of the GEBIK equations in BRTSim only required to write the kinetic reactions in Eq. (11) using the standard BRTSim format in the `Param.inp` file, while the PEST block in the `Param.inp` file was used with the PEST software (Doherty et al., 2016) to estimate the parameters $k_1$, $k_2$, $K_1$, $K_2$, $Y$, and initial biomass concentration $B_0$ (see Table 2). The GEBIF equations were implemented as a post-processing script. Figure 6 shows agreement between BRTSim modeling of bulk $NO_3^-$ and $N_2O$ concentrations, and $\delta^{15}N$-$NO_3^-$ against observations at 20 $^oC$ ($R^2 = 0.98$ and normalized root mean squared error NRMSE = 4.16%).

| $T$ | [$^oC$] | 20 |
|---|---|---|
| $2 \cdot k_1$ | [1/s] | 5.42×10$^{-4}$ |
| $k_2$ | [1/s] | 4.56 ×10$^{-4}$ |
| $K_1$ | [mol/L] | 2.723 |
| $K_2$ | [mol/L] | 2.309 |
| $Y$ | [mg/mol] | 295.7 |
| $B_0$ | [mg/L] | 1.073 |

Table 2
Summary of parameters used in the GEBIK and GEBIF equations describing experiments in Figure 6.

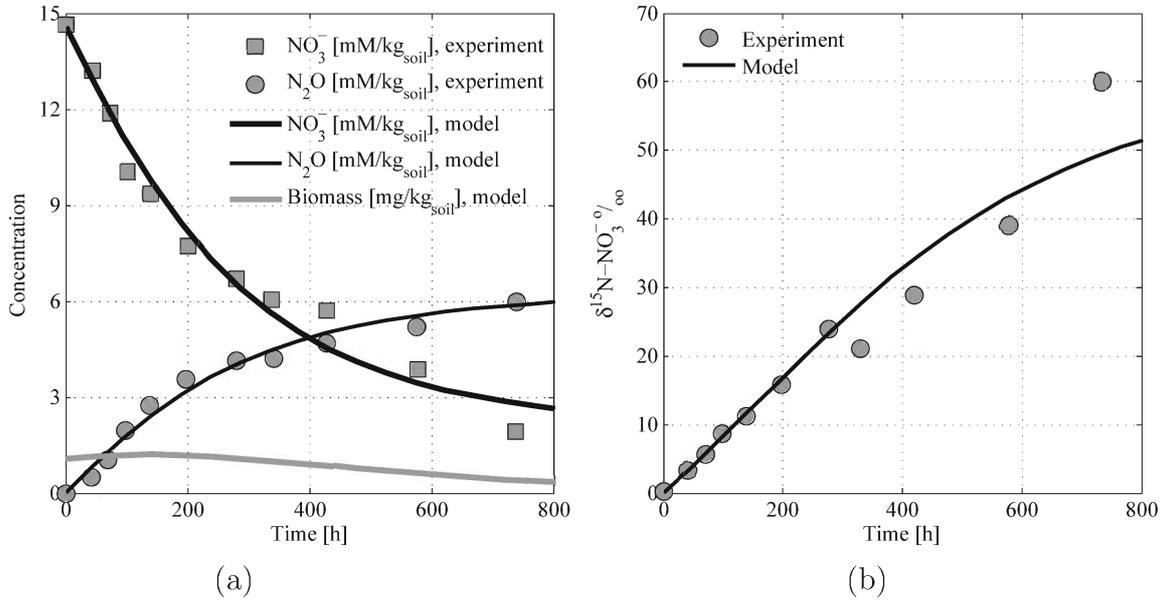

Fig. 6. Bulk $NO_3^-$ and $N_2O$ concentration as a function of time during $NO_3^-$ denitrification in a soil microcosm and corresponding $\delta^{15}N$-$NO_3^-$ at 20$^oC$. Experiments from Mariotti *et al.* (1981).

To analyse how $T$ affected microbial metabolism and the consequent kinetic isotopic fractionation, a number of simulations were run in BRTSim for $T$ ranging between 5 and 50 $^oC$ (Figure 7). The microbial temperature response was described in this case using Eq. (7) with $T_{LB} = 15$ $^oC$ and $T_{UB} = 40$ $^oC$, thus representing the response of typical soil mesophiles (Rittman and McCarty., 2001). The $\delta^{15}N$-$NO_3^-$ values at 800 h decreased substantially at high and low temperatures, when microbial activity reduced as compared to the optimal microbial activity corresponding to the plateau from about 25 to 32 °C for the given $T_{LB}$ and $T_{UB}$ values (Figure 7).

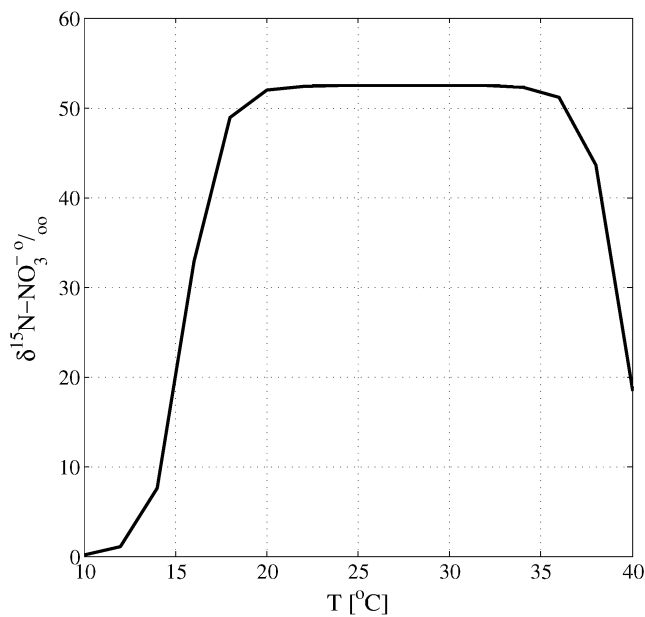

Fig. 7. Representation of $\delta^{15}$N-NO$_3^-$ at 800 h as a function of temperature.

*3.4 Case 4 - Dispersion and biodegradation of agrochemicals*

The capability to describe the dynamics of chemicals dispersion in the environment is becoming fundamental to develop predictive tools for scenario analyses and environmental risk assessments. This is fundamentally linked to the level of coupling of (eco)hydrological processes of water flow, biogeochemical degradation of reactive compounds, and the microbiological functions in the environment. BRTSim has recently been applied to a number of cases in which nutrients and agrochemical interactions play a key role in governing the soil and water quality. In those instances, the full BRTSim capabilities were used as compared to the examples presented above, and included multi-phase flows with complex boundary conditions and comprehensive reaction networks. For example, the urine biogeochemical decomposition pathways have been investigated for use in agriculture as an alternative to mineral fertilization (Tang and Maggi, 2016), and for recycling matter and nutrients in a bioregenerative soil-based life-support system for manned space missions (Maggi et al., 2018). Similarly, the nitrogen cycle coupled to suspended particle matter dynamics in aqueous ecosystems has been described to assess the water quality in streams, estuaries and coastal waters (Tang and Maggi, 2018), while atrazine and glyphosate dispersion and decomposition in croplands have been investigated to highlight the vadose zone and aquifer contamination under variable hydrometeorological and ecohydrological conditions in deterministic (la Cecilia and Maggi, 2017; la Cecilia et al., 2018) and stochastic uncertainty analyses (Porta et al., 2018).

Not all these examples are treated in detail here; the reader can access quoted works and gather all required information to replicate earlier results with the released BRTSim package available in Appendix. Rather, we only provide a general description and key features of the case of atrazine (ATZ) herbicide dispersion, mobility and biological degradation in a cropland, and uncertainty analyses. Here, BRTSim was used to lever in full the coupling capability to interlace hydrological, ecohydrological, and biogeochemical processes that govern water flow along the vertical soil profile, the concentration of biodegradation metabolites in aqueous solution and adsorbed to the mineral phase, and gas genesis and emissions to the atmosphere over 100 year time scale. An heterogeneous soil profile was described with 1 grid element as the atmosphere and 18 elements down to a depth of 5 m characterized by soil textural and hydraulic properties from a site located in West Wyalong (about 470 km West of Sydney) in New South Wales, Australia, where wheat is largely cultivated. Experimental hydrometeorological and ecohydrological conditions (daily precipitation and evapotranspiration) were used as the boundary conditions. Analysis of ATZ dynamics included a reaction network comprising 40 primary chemical species (including virtual species for molecular tracking), 6 microbial species, 4 secondary aqueous species, 20 mineral-adsorbed species in equilibrium with aqueous species, and 7 secondary gas species (Figure 8). Note that parameters were determined from individual experiments and integrated into the reaction network following the validation by construct approach in McCarl and Apland (1986). BRTSim was next used to analyse scenarios of ATZ application rates on the water quality in the root zone, the vadose zone and the aquifer, including the ATZ biodegradation half-time, the breakthrough curves, and the effect on soil quality including pH.

An animation of the reference simulation can be accessed at the link in Appendix along with the `Param.inp` file and various post-processing scripts written in Matlab 2011b available to the reader to replicate the results or modify inputs and produce new scenarios. The overall results of those analyses are presented in detail in la Cecilia and Maggi (2017) and Porta et al. (2018), and are summarized only briefly as follow: (i) about 97% of applied ATZ was biodegraded in the root zone, which is highly populated by ATZ oxidizers and hydrolizers; (ii) the beneath aquifer was contaminated by ATZ within 10 years, and contamination was persistent, with removal depending only on dilution and adsorption to the mineral phase; (iii) the ATZ half-life increased substantially with depths and ranged from 200 days in the top soil to 348 days or more from 1 m depth downwards; (iv) the greatest dependence of biodegradation reactions were related to the C source, which fluctuations resulted in a fluctuating content of ATZ oxidizers and, as a consequence, ATZ removal; and (v) the coupling between hydrological and biogeochemical processes had a long-term effect on predicted ATZ contamination variability.

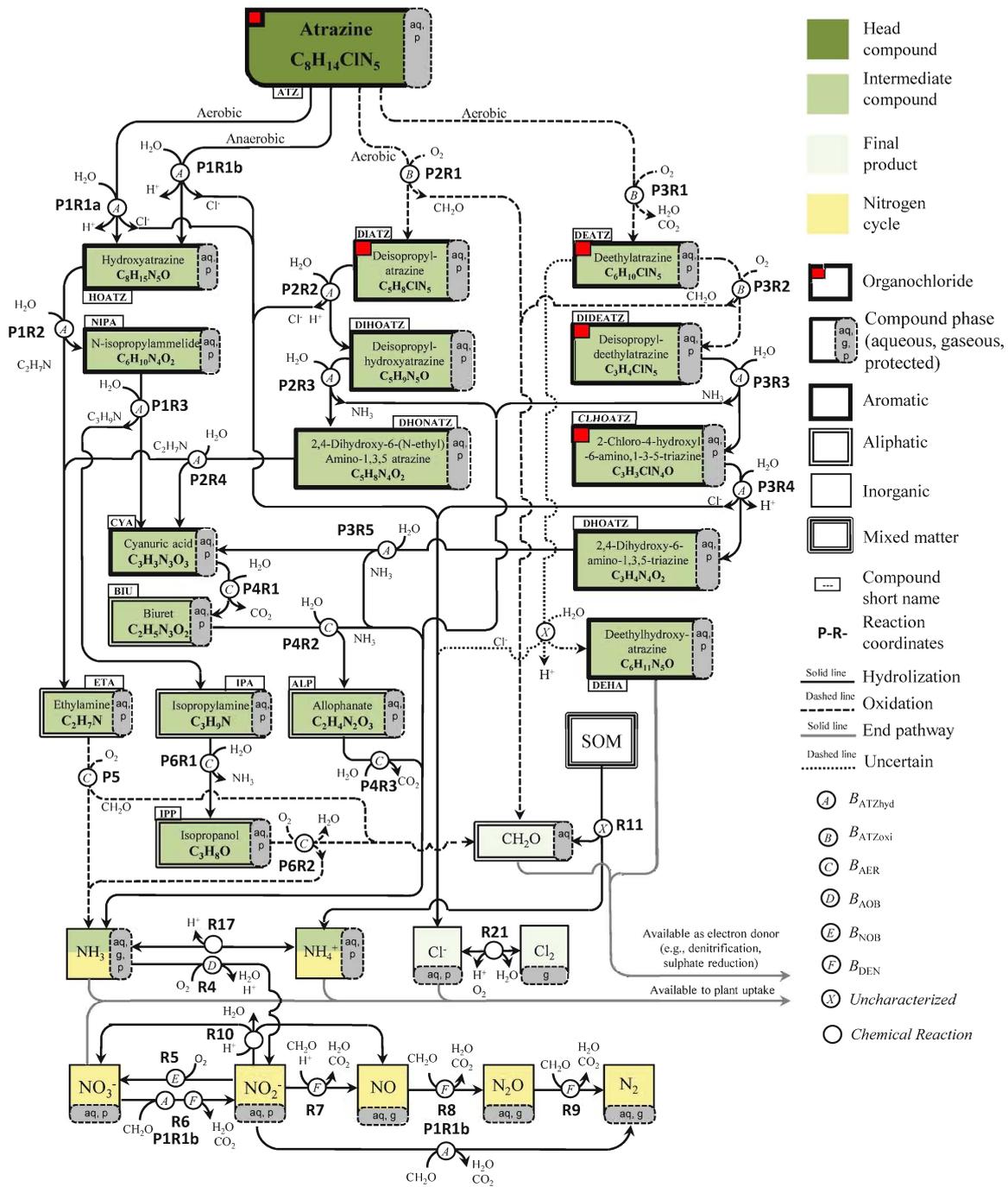

Fig. 8. Atrazine biodegradation reaction network and its intersection with the nitrogen cycle in soil. Reaction network redrawn from la Cecilia and Maggi (2017).

## 4 Discussion

Beside the relatively large number of codes existing for research purpose, only some solvers, such as those highlighted in the introduction, have reached a wider scientific audience and general use in specific hydrological and biogeochemical applications. These, however, remain highly special-purpose rather than general-purpose. The proposed computational concepts embedded in BRTSim aims at filling in this gap with general-purpose capabilities that make it flexible and applicable to coupled hydrological and biogeochemical problems. BRTSim bases on relatively simple explicit or combined explicit-implicit solvers that have largely been tested in earlier works against mass conservation of bulk phases and dissolved species, and on equilibrium reactions, as well as against analytical solutions to chemical and biochemical reactions. However, algorithms are under continuing development to improve numerical accuracy, application flexibility, and computational performance. Hence, BRTSim is also a platform where new algorithms are implemented and tested to improve older and less performing algorithms.

As compared to other solvers, BRTSim only uses one text input file (`Param.inp`) structured by blocks in ascii format. The only optional files that may be needed are the boundary flows provided in .TXT files called in `Param.inp` - generally, one file organized by columns will be enough to instruct BRTSim such as in Case 4 presented here. This organization of instructions to BRTSim is particularly simple, and increases readability and data manipulation by non-expert users or automated scripts.

It is to be noted that the integration of diverse processes in BRTSim, each one requiring time scales that may largely vary from one another, may result in the user to analyse, and likely find, best convergence criteria associated to a specific problem. For this reason, iteration number and tolerances can be set for the various processes but it is practically impossible to exhaustively make *a priori* recommendations on mutual compatibility across criteria applied to each process. The BRTSim framework is therefore designed to allow the user to correct potential instabilities, which depend on the specific problem to be solved, and set the numerical solvers to converge by criteria with an arbitrary level of accuracy.

BRTSim is an evolving tool that is foreseen to expand the current capabilities and increase the diversity of potential users. As compared to earlier versions of BRTSim, revised algorithms have given an enormous potentiality especially in terms of computational performance. The source code is written in the Matlab environment and levers the vectorized computational structure available in Matlab, as well as a number of functions optimized for performance. The advantage of using the executable of BRTSim is that the user can access the Matlab Runtime (MCR) freely available for any platform and run BRTSim without compatibility issues. Details on launching BRTSim are reported in the User Manual and Technical Guide available at the link in Appendix. BRTSim is currently free and a full licence is provided at the link to the download page in Appendix.

For the characteristics summarized above, BRTSim can provide extraordinary insights in soil physics and biogeochemistry, soil and water quality assessment and prediction, soil bio- and phyto-remediation estimation, and in all environmental research contexts where physical, chemical, and biological processes co-exist and affect each other with complex nonlinear feedbacks with water flow. Planned new implementations include application of solvers to an explicit two-dimensional domain, and links to web service providers for use with real-time hydrometeorological conditions in environmental applications among others.

# 5 Conclusion

Coupled hydrological and biogeochemical computational principles in BRTSim have briefly been described and four applications have been demonstrated in the description of the aquifer, vadoze zone, and water table dynamics (Case 1), biological clogging of a soil column (Case 2), consumption and isotopic fractionation of a substrate by a soil microbial population (Case 3), and a comprehensive modeling of agrochemical dispersion and biodegradation in croplands (Case 4). These cases, although exemplificative, occur in the watershed with feedbacks that involve both hydrology and biogeochemistry and, hence, results highlight the importance to ultimately integrate to a great detail the existing knowledge of catchment-specific processes involving both the abiotic and biotic worlds. BRTSim blends into a single tool a number of capabilities of advanced models with new general-purpose features to conduct analytical inspections, explore ecosystem scenarios responses, test hypothesis and validate approaches, and make predictions in a wide range of environmental science and engineering problems.

# 6 Appendix

The BRTSim package can be downloaded from the hyperlink https://www.dropbox.com/sh/wrfspx9f1dvuspr/AAD5iA9PsteX3ygAJxQDxAy9a?dl=0, wich includes a version history, and the BRTSim v3.1a compressed folder with executable, licence.txt file, User Manual and Technique Guide, and a number of examples. The licence file has limited time validity but is updated regularly to allow continuing use.

BRTSim is compiled for Win64 operating systems using Matlab 2011b, and for GLNXA64 using Matlab 2017a. Hence, the user may need to download the MCR version 7.16 (for Win64) and MCR 9.2 (for GLNXA64) at the same link reported above.

# 7 Acknowledgment

Many thanks to Fiona H.M Tang and Daniele la Cecilia for comprehensive use and testing of beta versions of BRTSim during the course of development and improvements since its first version. The author was partly supported by the Sydney Research Excellence Initiative (SREI 2020) EnviroSphere program, the Mid Career Research Award (MCR), and the Sydney Research Accelerator Fellowship (SOAR) of The University of Sydney.

# References

Abbott M.B., Bathurst J.C., Cunge J.A., O'connell P.E. and Rasmussen J., (1986), An introduction to the European Hydrological SystemSysteme Hydrologique Europeen,SHE, 2: Structure of a physically-based, distributed modelling system, Journal of hydrology, 87(1-2), pp.61-77.

Allen R.G., Pereira L.S., Raes D., and Smith M., (1998), FAO - Food and Agriculture Organization of the United Nations, Natural Resources and Environment, Rome, 1998. ISBN 92-5-104219-5. http://www.fao.org/land-water/databases-and-software/eto-calculator/en/2

Beven K., (1989), Changing ideas in hydrologythe case of physically-based models. Journal of hydrology, 105(1-2), pp.157-172.

Beven K., and Freer J. ( 2001), A dynamic topmodel. Hydrological processes, 15(10), pp.1993-2011.

Brangari, A.C., Fernndez-Garcia, D., Sanchez-Vila, X. and Manzoni, S., (2017), Ecological and soil hydraulic implications of microbial responses to stress-A modeling analysis. Advances in Water Resources.

Brooks RH and Corey AT, (1962), Hydraulic properties of porous media, Hydrology Papers, A14407-599281, Colorado State University, Fort Collins, CO, USA.

Brutsaert W. (2000), A concise parameterization of the hydraulic conductivity of unsaturated sopils, *Wat. Res. Res* 23, 811-815.

La Cecilia D. and F. Maggi, (2017), In-situ atrazine biodegradation dynamics in wheat (Triticum) crops under variable hydrologic regime, Journal of Contaminant Hydrology 203 (2017) 104121.

La Cecilia D., Tang F., Coleman N.V., Conoley C., Vervoort R.W., and Maggi F., (2018), Glyphosate dispersion, degradation, and aquifer contamination in vineyards and wheat crops in the Po Valley, Italy, Journal of Agricultureal and Food Chemistry, under review.

Chiari G., Genovesi R., Raimondi S., Sarno G., and Tarocco P., (2016), database accessed on 01.02.2017 at http://cloud.consorziocer.it/FaldaNET/retefalda/index. CG and GR from Consorzio di Bonifca di secondo grado per il Canale Emiliano-Romagnolo; RS from Coop. I.ter; SG from Regione Emilia-Romagna. Servizio ricerca, innovazione e promozione del sistema agroalimentare; TP from Regione Emilia-Romagna. Servizio Geologico, Sismico e dei Suoli.

Coleman K., Jenkinson D.S., (1996), RothC-26.3 - A model for the turnover of carbon in soil. In: Powlson, D.S., Smith, P., Smith, J.U. (Eds.), Evaluation of Soil Organic Matter Models Using Existing Long-Term Datasets. Springer-Verlag, Heidelberg, pp. 237-246.

Cosby BJ ,Hornberger GM, Clapp RB Ginn TR, (1984), A Statistical Exploration of the Relationships of Soil Moisture Characteristics to the Physical Properties of Soils, Water Resources Research 20(6) 682690.

Cunningham A.B., Sharp R.R., Hiebert R., and James G., (2003), Subsurface biofilm barriers for the containment and remediation of contaminated groundwater. Bioremediation Journal, 7(3-4), 151-164.

Del Grosso S.J., Parton W.J., Mosier A.R., Ojima D.S., Kulmala A.E., and Phongpan S.,

(2000), General model for N2O and N2 gas emissions from soils due to dentrification, Global Biogeochemical Cycles 14, 1045-1060.

Doherty J., C. Muffels, J. Rumbaugh, M. Tonkin, (2016), PEST Model-Indipendent Parameter Estimation & Uncertainty Analysis, http://www.pesthomepage.org/PEST.php.

Yan Z., Bond-Lamberty B., Todd-Brown K.E., Bailey V.L., Li S., Liu C., and Liu C, (2018), A moisture function of soil heterotrophic respiration that incorporates microscale processes. Nature communications, 9(1), 2562.

Yang S.F. and Li X.Y., (2009), Influences of extracellular polymeric substances (EPS) on the characteristics of activated sludge under non-steady-state conditions. Process Biochemistry, 44(1), pp.91-96.

Feller G., (2010), Protein stability and enzyme activity at extreme biological temperatures. Journal of Physics: Condensed Matter, 22(32), 323101.

Van Genuchten MT, (1980), A Closed-form Equation for Predicting the Hydraulic Conductivity of Unsaturated Soils, Soil Sci. Soc. Am. J. 44, 892-898.

Gharasoo M., Thullner M., Elsner M., (2017), Introduction of a new platform for parameter estimation of kinetically complex environmental systems, Environmental Modelling & Software, 98, 12-20.

Heinen M., (2006), Simplified denitrification models: overview and properties. Geoderma 133(3-4), 444-463.

Johnson K.A., Simpson Z.B., Blom T., (2009), Global Kinetic Explorer: a new computer program for dynamic simulation and fitting of kinetic data. Anal. Biochem. 387 (1), 20-29.

Johnson C.D., and Truex M.J.M.J., (2006), RT3D reaction Modules for natural and enhanced attenuation of chloroethanes, chloroethens, chloromethanes, and daughter products, Pacific Northweest National Laboratory, PNNL-15938.

Keller EF, and Segel LA (1971) Traveling bands of chemotactic bacteria: a theoretical analysis, J. Theor. Biol. 30:235.

Li C., Frolking S., and Frolking T.A., (1992), A model of nitrous oxide evolution from soil driven by rainfall events: 1. Model structure and sensitivity, Journal of Geophysical Research 97, 9759-9776.

Liang X., Lettenmaier D.P., Wood E.F., Burges S.J., (1994), A simple hydrologically based model of land-surface water and energy fluxes for general-circulation models. J. Geophys. Res. -Atmos 99, 14415-14428.

Michaelis L. and Menten M., (1913), The Kinetics of Invertase Action, Biochem. Z. 49, 333, English translation by R.S. Goody and K.H. Johnson in Biochemistry 2011, 50, 8264-8269.

Maggi F., and la Cecilia D., (2016) Implicit Analytic Solution of MichaelisMentenMonod Kinetics. ACS Omega, 1(5), 894-898.

Maggi F. and Porporato A., (2007), Coupled moisture and microbial dynamics in unsaturated soils. Water resources research, 43(7).

Maggi, F., and Riley ,W.J. (2009) Transient competitive complexation in biological kinetic isotope fractionation explains non-steady isotopic effects: Theory and application to denitrification in soils. J.Geophys. Res.- Bio. 2009, 114, G04012. DOI: 10.1029/2008JG000878.

Maggi F., and Riley, W.J. (2010) Mathematical treatment of isotopologue and isotopomer speciation and fractionation in biochemical kinetics. Geochim. Cosmochim. Acta 2010, 74,

1823.

Maggi F., Tang F.H.M., and W.J. Riley, (2018), The thermodynamic link between substrate, enzyme, and microbial dynamics in Michaelis-Menten-Monod kinetics, International Journal of Chemical Kinetics, DOI 10.1002/kin.21163.

Maggi F., Tang F.H.M., C. Pallud and C. Gu, (2018). A urine-fuelled soil-based bioregenerative life support system for long-term and long-distance manned space missions. Life Sciences in Space Research 17, 1-14.

Manzoni S. and Porporato A., (2009), Soil carbon and nitrogen mineralization: theory and models across scales. Soil Biology and Biochemistry 41(7), 1355-1379.

Mariotti A., J.C. Germon, P. Hubert, P. Kaiser, R. Letolle, A. Tardieux, P. Tardieux, (1981), Experimental determination of nitrogen kinetic isotope fractionation - Some principles - Illustration for the denitrification and nitrification processes, Plant and Soil 62(3), 413-430.

McCarl B.A., and Apland J., (1986), Validation of linear programming models. Southern Journal 621 of Agricultural Economics, 18, 155164.

Monod J., (1949), The growth of bacterial cultures, Annu. Rev. Microbial 3, 371-394.

Mualem Y. (1976), A new model for predicting the hydraulic conductivity of unsaturated porous media, *Water Res. Res.* 12(3), 513-522.

Mualem Y. and G. Dagan (1978), Hydraulic conductivity of soils: unified approach to the statistical models, *Soil Science Society of America Journal* 42(3), 392-395.

Neitsch S.L., Arnold J.G., Kiniry J.R., Srinivasan R., Williams J.R., (2000), Soil and Water Assessment Tool User's Manual, Published 2002 by Texas Water Resources Institute, College Station, Texas TWRI Report TR-192

Parton W.J., A.R. Mosier, D.S. Ojima, D.W. Valentine, D.S. Schimel, K. Weier, A.E. Kulmala, (1996), Generalized model for N2 and N2O production from nitrification and denitrification, Global Biogeochemical Cycles 10(3), 401-412.

Porta G., la Cecilia D., Guadagnini A., Maggi F., (2018), Implications of uncertain bioreactive parameters on a complex reaction network of atrazine biodegradation in soil, Advances in Water Research (in press).

Potter C.S., Riley R.H., Klooster S.A., (1997), Simulation modeling of nitrogen trace gas emissions along an age gradient of tropical forest soils, Ecological Modelling 97, 179-196.

Rittmann B.E. and P.L. McCarty, (2001), Environmental Biotechnology: Principles and Applications, McGraw-Hill,pp. 754.

Ross N., Villemur R., Deschnes L., and Samson R., (2001), Clogging of a limestone fracture by stimulating groundwater microbes. Water Research, 35(8), 2029-2037.

Skopp J., Jawson M.D., and Doran J.W, (1990), Steady-state aerobic microbial activity as a function of soil water content. Soil Science Society of America Journal, 54(6), 1619-1625.

Tang F.H.M., and Maggi F., (2016). Breakdown, uptake and losses of human urine chemical compounds in barley (Hordeum vulgare) and soybean (Glycine max) agricultural plots, Nutrient Cycling in Agroecosystems, DOI 10.1007/s10705-016-9768-z.

Tang F.H.M. and Maggi F., (2018), Biomodulation of nitrogen cycle in sedimentary habitat, Journal of Geophysical Research Biogeosciences, DOI: 10.1002/2017JG004165.

Thullner M., Kampara M., Richnow H.H., Harms H., Wick L.Y., (2008), Impact of bioavailability restrictions on microbially induced stable isotope fractionation. 1. theoretical calculation.

Environ. Sci. Technol. 42 (17), 6544-6551.

Yu C., and Zheng CC., (2010), HYDRUS: software for flow and transport modeling in variably saturated media, Ground Water 48(6), 787-791.

Xu T., Spycher N., Sonnenthal E., Zhang G., L. Zheng, Pruess K., (2011), TOUGHREACT version 2.0: a simulator for subsurface reactive transport under non-isothermal multiphase flow conditions, Ground Water 37(6), 763-774.

Well R., I. Kurnegova, V. Lopes de Gerenyu and H. Flessa, (2006), Isotopomer signatures of soil-emitted $N_2O$ under different moisture conditions - A microcosm study with arable loess soil, Soil Biology & Biochemistry 38, 2923-2933.

MUSIC, accessed at https://ewater.org.au/products/music/ on 12th April 2018.

SGSS (2016), database accessed on 01.02.2017 at `https://applicazioni.regione.emilia-romagna.it/cartografia_sgss/user/viewer.jsp?service=pedologia&bookmark=1%22`. Regione Emilia-Romagna. Servizio Geologico, Sismico e dei Suoli.